# A Theoretical Limit to Physicalism: A Non-Technical Explanation of the Gemini Theorem

Catherine M Reason[1]

*Independent Researcher*

*London, United Kingdom*

*The Gemini theorem asserts that, given certain reasonable assumptions, no physical system can be certainly aware of its own existence. The theorem can be proved algorithmically, but the proof of this theorem is somewhat obscure, and there exists very little literature on it. The purpose of this article is to provide a brief non-technical summary of the theorem and its proof, with a view to stimulating critical discussion of the proof and its implications. Since the theorem implies that a violation of the conservation of energy will take place within the brains of conscious human beings, it has obvious implications for any physical theory.*

**The Basis of the Theorem**

The purpose of this article is to provide readers with a simple, non-technical explanation of the Gemini theorem. The theorem, first outlined by Caplain (1995, 2000)[2] and developed by Reason (2016, 2018) establishes that no physical system possessing human-level intelligence can ever be certain of possessing conscious awareness. This theorem depends on a small number of well-defined assumptions and operationalized definitions, and has no connection with the philosophy of mind. Although the theorem refers to the mind-body problem, the proof of the theorem is mathematical rather than philosophical, and philosophically minded readers should bear in mind that a proof of a mathematical theorem is quite different from an argument defending a philosophical thesis. A useful explanation of the difference can be found in a paper by Terence Parsons (1996).

This article is divided into three sections. The first describes the Gemini theorem itself. The second part analyzes objections which have been made frequently by both philosophers and psychologists. Finally I shall make some concluding remarks about the lack of attention which has been given to this type of reasoning over the last fifty years.

The proof of the Gemini theorem is extremely simple, and could be scribbled on the back of an envelope. However it is a mathematical proof, not a philosophical argument, and therefore lies outside of the comfort zone of most psychologists and philosophers of mind. The basis of the theorem must therefore be explained with particular care. Philosophers who find themselves registering objections at various stages of the proof are requested to withhold their objections until they have read the subsequent section, since it is likely they will find the answers to their difficulties there.

---

[1]Correspondence to: CMRneuro@gmail.com

[2]Caplain's first paper has been criticized by Bojadziev (2000) and in a review by Dunlop (2000). Caplain's own 2000 paper is a response to these criticisms.

A key feature of the Gemini theorem is that it does not require any assumptions to be made about the nature of consciousness itself. It relies instead on an operational property possessed by all healthy conscious human beings. This is the property of being able to assert that they do, in some sense, exist -- a property which I have elsewhere referred to as *self-certainty* (Reason 2016). This property can be expressed as the ability to answer YES to questions such as "Can I be certain that I am conscious?" or "Can I be certain that I have at least the illusion that I exist?" Such an ability can be regarded as a specific element of a general set, which comprises the ability to answer YES to any question.[3]

Now it is clearly the case that, if it can be found that some particular type of system is unable to answer YES to any question in the set, then such a system must be incapable of self-certainty. Since we are interested in the question of whether human consciousness can be wholly physical, we will restrict our considerations to systems which have humanlike powers of intelligence and reasoning. Such a power must, of course, include the ability to use both arithmetic and classical logic. So, if we can show that no physical system using arithmetic and classical logic can answer YES to any YES/NO question with certainty, then we can show that human consciousness cannot be entirely physical.

Following the definition given in Reason (2018), let us explicate that a physical system is any system entirely made up of, or supervening on, a set of objectively real properties, such that the system has the capability to evolve in time. This assumption we shall call the *principle of physicalism*.

It is obviously the case that in such a physical system, any function -- such as the function of finding the answer to some YES/NO question -- will supervene on some physical process. Of course, physical processes will differ in their ability to perform such a function. Some will do so correctly, others incorrectly or not at all. We shall express this notion as the *axiom of fallibility* -- for any given function, any given process may perform that function either correctly or incorrectly. In order to know which, one has to determine something about the process.

Here we note that the function of finding out whether some process is correct, can itself be represented as answering a YES/NO question -- such as "Does this process perform this function correctly". From the principle of physicalism, this function must be performed ultimately by some physical process -- which will itself be subject to the axiom of fallibility.

We can express this situation in terms of a couplet of sentences:

---

[3] If these questions are assumed to be expressed in the English language, then the set of such questions is of course recursively enumerable.

The function of answering a YES/NO question must, by the principle of physicalism, be performed by some physical process.

The correctness of any physical process must, by the axiom of fallibility, be established by answering some new YES/NO question.

This is the *Gemini Couplet*, and quite clearly it induces a non-terminating sequence of processes -- which is to say, an infinite regress.  It is therefore obvious that, for any physical system capable of reasoning in classical logic with arithmetic, no YES/NO question can be answered with certainty.  Since the set of YES/NO questions includes self-certainty, it is obvious that any physical system capable of reasoning using classical logic with arithmetic induction is incapable of self-certainty.  We can therefore express the Gemini theorem as follows:

"No physical system capable of humanlike reasoning can decide any YES/NO question with absolute certainty in a finite number of steps."

This is the "back-of-an-envelope" version of the proof -- the formal proof given in Reason (2018) is much longer, but amounts to the same thing.  However the proof can also be expressed as an effective procedure in terms of the following *certainty lemma*[4].  Here we assume that M is some physical system capable of humanlike reasoning (including the ability to use arithmetic and classical logic):

We start by assuming that M is certain of some proposition p, which can be represented as the answer to some YES/NO question.  By the principle of physicalism, the process of determining p must supervene on some physical process, say X.  From the axiom of fallibility, M can deduce that X may be fallible.  Since M can reason using classical logic, M can deduce that if X is fallible, then its answer to the YES/NO question corresponding to p may be incorrect, in which case M cannot be certain of p.  Therefore to be certain of p, M must have some way of determining that X is operating correctly.  By the principle of physicalism, this must supervene on some physical process, say X*.  But what process is X*?  It cannot be X, since M relies on the process X* to ensure that X is operating correctly.  X therefore depends on X*, not the other way round.

Therefore M must assume that X* is some new process X', which is different from X.  But from the axiom of fallibility, M can deduce that X' may be fallible.  Since M's certainty of p depends on X, and the correctness of X depends on X', M must have some means of determining that X' is operating correctly.  Let us call this process X**.  But what process is X**?  It cannot be X', since the correctness of X' depends on the correctness of X**.  Neither can it be X, since establishing the correctness of X requires the correctness of X'.  Therefore we require some new process X" ... and so on ad infinitum.  This leaves us with no choice but to discard the assumption which led to the regress -- that is, the assumption that M can be certain of p.

The purpose of this lemma is to illustrate that it is unnecessary to make assumptions about the nature of certainty in order to prove the Gemini theorem.  The assumption that certainty in a physical system is possible is in itself sufficient to generate an infinite regress in any humanlike physical system using both arithmetic and classical logic.  The certainty lemma can be regarded as an algorithm which takes as its input some YES/NO question, and decides whether or not that question can be answered in a finite number of steps with certainty .  For any physical system capable of humanlike reasoning, the algorithm will always return the answer that no YES/NO question can be answered in this way[5].  Because the proof of the Gemini theorem can be expressed algorithmically in this way, it cannot be regarded as a philosophical argument but must be understood as a mathematical proof.

---

[4]This is a slightly modified version of the certainty lemma from Reason (2018).

**Objections which have been raised against the Gemini theorem and its implications**

This list includes all the objections which I have received to date.  Some of these are covered in more detail in my previous paper (Reason 2018); others are listed here for the first time.  They are collected here together for simplicity and convenience.  All are based on fairly straightforward misunderstandings of the theorem, its proof, and its application.

*The requirement for proving an infinite sequence of propositions is not a necessary epistemological assumption.*

This is correct, but the requirement for such an infinite sequence does not follow from any epistemological assumptions about the nature of certainty.  It follows simply from three assumptions:  Firstly, that any humanlike physical system must be capable of humanlike reasoning;  secondly, the principle of physicalism; and thirdly, the axiom of fallibility.  Any physical system capable of humanlike reasoning can prove that its certainty depends on being able to prove an infinite sequence of propositions; it is not necessary to assume this, and this is shown in the above certainty lemma.  In any system capable of humanlike reasoning, the Gemini couplet is itself sufficient to generate such an infinite sequence.

*M can be certain of some proposition because it has access to additional evidence which is not available in the possible world where the physical process* X *fails.*

This is incorrect, and is based on a failure to appreciate that the Gemini proof applies to all YES/NO questions.  This includes questions of the form "Does this additional evidence exist?".  Any means used by M to determine the existence of this additional evidence, must by the principle of physicalism supervene on some physical process.  This process can be regarded as performing the function of answering a question of the form "Does this evidence exist?"  Clearly, the Gemini theorem will apply to any such process.

---

[5]Strictly one might argue that the certainty lemma is not really a lemma, since it is really just the Gemini theorem proved by another means.  However to ensure that the Gemini theorem covers self-referential statements, it is necessary to prove an additional *Cartesian lemma*, which is described later in this article.  The certainty lemma and the Cartesian lemma together can be taken as providing a comprehensive proof of the Gemini theorem.

*It is quite possible to program a machine to respond "yes" to the question "Am I conscious?" without having any idea what it is doing.*

Once again this is correct, but it does not provide a way round the Gemini theorem. Since it is required that any humanlike physical system must be capable of humanlike reasoning, such a system can itself deduce that such a program may exist. Needless to say, the axiom of fallibility will apply to any such program, and this the system will also be able to deduce. The Gemini couplet therefore applies as before.[6]

*Since the definition of self-certainty is completely operationalized, there is no basis for interpreting statements such as "I exist" as being either true or false.*

The physical systems covered by the Gemini theorem are required to be capable of humanlike reasoning, which means they must be capable of treating the answers to all YES/NO questions as statements which can be either true or false. The inability of such a physical system to perform the operation of self-certainty then follows naturally; it is not necessary that the operation itself be interpreted in any way by any external observer.

This is just equivalent to saying that any physical system covered by the Gemini theorem, must be capable of reasoning according to some logical system which supports an interpretation in terms of truth values. This is a necessary precondition for any system to reason according to the principles of classical logic. Since the Gemini theorem covers all YES/NO questions, there is no need to consider the meanings of the individual YES/NO questions themselves.

*There is no theoretical reason why even human beings should actually be capable of self-certainty.*

This response comes most often from psychologists and cognitive scientists, and is absolutely correct. The ability of human beings to perform self-certainty is an empirical matter -- it cannot be established by any sort of theoretical argument. This observation invites two responses. The quick response is that if self-certainty is possible in humans, then this entails a clear empirical prediction, which is the $\chi$ effect described in my previous papers. Since this prediction can in principle be tested, no further theoretical discussion of this point is strictly necessary.

The more detailed response, however, is that it is not clear to me that all of those who claim that self-certainty is not possible in human beings have really thought all that much about what they are committing themselves to. It is fairly easy to argue that human beings cannot be certain that they have conscious existence if one has implicitly in mind some particular theoretical or philosophical notion of what consciousness is. But the implication of the Gemini theorem is that self-certainty is impossible in any physical system *regardless* of what consciousness is assumed or understood to be. It is impossible, according to the Gemini theorem, for a physical system to be certain even that it possesses the *illusion* of consciousness. Indeed it would be impossible for a physical system to be certain of its own existence in any sense. Those who claim that self-certainty is impossible in human beings are therefore committing themselves to the notion that they cannot even be sure they have any form of existence whatsoever. They cannot even be certain that they are not dead (in the sense of being completely non-existent, as opposed to preserved in some sort of virtual reality afterlife). Those who make this claim should therefore be quite explicit that they do, indeed, intend this commitment.

---

[6]More specifically, the axiom of fallibility will apply to any process which correlates a state of the system with the truth-value of some proposition.

*The theorem relies in a two-valued logic, and may not work for a multivalued logic.*

This is not entirely correct. By expressing propositions in terms of *certain* versus *uncertain*, one can actually transform all the required propositions into YES/NO questions -- that is, answers to questions of the form "Can I be certain of some proposition p?" Only if one is prepared to endorse logical systems in which propositions can be both certain and uncertain, or neither certain nor uncertain, can the Gemini theorem be avoided. It is not clear, at least to this author, what sort of meaningful interpretation such a logic could support.

*The proof only applies if one assumes a foundationalist epistemology*

This is primarily a criticism of Caplain's original papers, which were expressed in fairly traditional philosophical language and were therefore vulnerable to a considerable amount of equivocation over terms. The criticism itself results from a simple misunderstanding of the theorem. The theorem itself does not assume or imply that self-certainty, or indeed any other kind of certainty, is possible in human beings. The possibility of self-certainty in human beings can only be established empirically, and it applies only to a specific limited class of propositions -- those that relate to consciousness or existence.

*Human beings do not need to "prove" that they are conscious or that they exist -- the fact that they are capable of asking the question "Am I conscious" already establishes that.*

We must not assume a priori that physical systems, or machines, possess some competency just because we ourselves are aware of possessing it. The objection here is evidently a version of the Cartesian "I think therefore I am" and is subject to the following lemma, adapted from Reason (2018):

A system reasoning using classical logic could get as far as asking "Do I exist?" and then argue, along the lines of Descartes' *Cogito ergo sum*, that it does not really matter if the physical process which subserves the answering of that question is accurate or not -- the mere fact that such a physical process exists is enough to answer the question in the affirmative. One can express this in the form of a *Cartesian Principle*: If M is asking "Do I exist?" then M must exist.

Descartes' *Cogito* can thus be represented as a deduction of the form:

If M is asking "Do I exist?" then M exists;

M is asking "Do I exist?";

Therefore M exists.

However this syllogism requires M to establish that it is the case that M is asking "Do I exist?". Since this clearly entails answering a YES/NO question, the Gemini proof will apply to it, as before.

*The same physical state may represent both the answer to some question and the question itself, so it is impossible even in principle for that answer to be wrong.*

This is essentially the philosophical theory known as *constitutivism* (see for example Shoemaker, 1990). In the context of the Gemini theorem it implies that, for example, the same physical state represents both the state of consciousness itself, and the answer YES to the question "Can I be certain that I am conscious?". Alternatively, both the state of consciousness and the answer YES to the such a question both result from the same physical process. According to the theory, therefore, it is not possible for a system to be in a state of having the answer YES to the question "Can I be certain that I am conscious?" unless it is, in fact, conscious.

However, as an escape route from the Gemini theorem, this simply will not work. To see why, we can express the constitutivist's reasoning as follows:

1  If a system of type M has property A, then it must also have property B;

2  A system $M_1$ of type M has property A;

3  Therefore $M_1$ has property B.

Clearly 2 can be expressed as the answer to a YES/NO question. Applying the Gemini couplet to this question must therefore generate an infinite regress. Constitutivism therefore offers no escape from the Gemini theorem.

*There is no obvious reason why the axiom of fallibility should be accepted.*

It is useful to assume the axiom of fallibility in order to express the Gemini proof as a simple couplet. But as was demonstrated in Reason (2018), the axiom of fallibility is not strictly necessary to prove the theorem. For any given YES/NO question, there will always be at least one physical process which will answer that question incorrectly[7]. One can therefore ask the question Q:

"Can I be certain that some physical process X answers some given question correctly?"

Since Q is clearly a YES/NO question, it must by the principle of physicalism be answered by some physical process, say X'. One can then ask the question Q':

"Can I be certain that the physical process X' answers question Q correctly?"

Since Q' is clearly a YES/NO question, it must by the principle of physicalism be answered by some physical process, say X". One can then ask the question Q":

"Can I be certain that the physical process X" answers the question Q' correctly?"

Once again this clearly induces an infinite regress.

*There is no reason in principle why a single physical process should not perform an infinite number of functions simultaneously.*

Both the detailed proof and the certainty lemma do indeed implicitly assume that every function is performed by a separate process. It may indeed seem reasonable, on the face of things, that allowing a single process to perform an infinite set of functions would rescue one from the consequences of the Gemini theorem. Unfortunately this is not the case. One could simply ask the question:

"Can I be certain, that it is not the case that every single function in the infinite set has been performed incorrectly?"

Such a question leads us once again into an infinite regress.

*The definition of a physical system is so general that it encompasses every conceivable type of system. There can therefore be no system to which the Gemini theorem does not apply.*

This is simply incorrect. The easiest way of seeing this is in terms of the certainty lemma above. Note that in the certainty lemma, the principle of physicalism is used to infer the existence of some process X, to which the axiom of fallibility is then applied. However if the principle of physicalism is not assumed, then no process X can be inferred. There is therefore no process to which the axiom of fallibility can be applied, and the certainty lemma does not apply.

*The Gemini theorem cannot be applied to statements about consciousness because the theorem tells us nothing about the nature of consciousness itself.*

---

[7]As was pointed out in Reason (2018), such a process can always be constructed by applying a logical NOT operator to the function which any given question.

Although the Gemini theorem explicitly requires us to make no assumptions at all about the nature of consciousness, people persist in raising this as an objection. The Gemini theorem establishes only that there is an inconsistency between self-certainty and the properties of certain types of physical system (those which are capable of humanlike reasoning). It is not intended, and is not required, to establish anything else about consciousness. This type of objection seems to illustrate most clearly the difference between a mathematical proof and a philosophical argument, and the apparent reluctance of many philosophers of mind to appreciate the difference.

*The assumptions on which the proof depends are ill-defined or incorrect.*

The Gemini proof depends on three key assumptions: Firstly, that M is capable of humanlike reasoning; secondly, the principle of physicalism; thirdly the axiom of fallibility. As was shown above, one can even dispense with the axiom of fallibility if one really wishes to do so. No correspondent has yet stated which of these assumptions is supposed to be unclear or incorrect.

*The proof is too simple and obvious to be correct. If the theorem were correct then it would already be well-known.*

This is undoubtedly my favorite of all the objections I have received to date. I suspect that the theorem may well have been discovered already on many occasions, but the discoverers concluded that the proof was far too simple and obvious to be correct, and that if correct it would already be well-known. They therefore neglected to publish their results.

**Discussion**

As we have seen, the proof of the Gemini theorem is extremely simple. Indeed I can see no reason why it should not have been discovered and proved by anyone at any time in the last sixty years. Yet I can find no reference to the theorem, or to any argument equivalent to it, anywhere in the philosophical literature at any time prior to Caplain's publication in 1995. This seems extraordinary. How can one account for such an omission?

Part of the answer may lie in the nature of philosophical argument itself. As Parsons (1996) has pointed out, philosophical arguments are constructed quite differently from mathematical or logical proofs. A proof is a statement derived from an explicit set of well-defined assumptions by means of clear rules of inference. An argument, by contrast, is a rationalization constructed to defend a pre-existing thesis, and is usually expressed in language which is intrinsically equivocal. Most, if not all, of the objections raised against the Gemini theorem in the preceding section can be attributed to misconceptions arising from the failure to appreciate the difference between a mathematical proof and a philosophical argument.

Considered as a mathematical proof, there are three ways one could attempt to challenge the basis of the Gemini proof. One could argue that certain of the assumptions on which it depends are untrue. Alternatively one could argue that it depends on implicit assumptions which are not explicitly stated. Or finally one could argue that the rules of inference are not correctly applied in the derivation.

None of the objections raised against the Gemini proof to date have adopted any of these approaches. Instead, all objections to date have fallen into one of three classes. Firstly, there are objections which simply fail to take account of all the assumptions and conditions laid out in Reason (2018). Secondly, there are objections which attempt to equivocate over the meanings of terms without regard to the logical relationships between them. And thirdly, there are objections which just ignore the proof entirely and simply express disapproval of its conclusion. (Since objections of this last type are not properly arguments at all, they are not included in the above list.)

Some correspondents have claimed a potential similarity between the Gemini theorem and the Lucas-Penrose argument which uses Godel's theorem to imply an intrinsic difference between the properties of machines and the capabilities of human minds. I believe this similarity is more apparent than real. Whereas the Lucas-Penrose thesis concentrates on a supposed difference between minds and machines, or between minds and formal systems, the Gemini theorem is actually concerned with the difference between machines or physical systems on the one hand, and the abstract logical systems they supposedly instantiate on the other (Indeed the "nature" of mind plays no part in the Gemini proof, since the concept of mind is fully operationalized from the outset.) It is this difference which is formalized as the axiom of fallibility. A physical system may be a perfect instantiation of some formal system, but that does not mean it can be certain of that fact. When one is dealing with actual physical entities, as opposed to abstract systems, then the issue of *epistemic possibility* arises. A proposition that cannot be true in any possible world may nonetheless *appear* to be possible from the viewpoint of some observer with imperfect knowledge. For example, one could argue that there is no possible world in which Fermat's last theorem is untrue. But a typical mathematician of the nineteenth century could not have been certain of this; for such a person, it would have been epistemically possible for Fermat's last theorem to be false. Similarly, propositions which are ontologically impossible in a formal system may nonetheless be epistemically possible to a physical instantiation of that system. This is the point upon which the Gemini theorem hinges.

The upshot is that the Gemini theorem does not depend on any assumptions about Godel's Incompleteness theorems. Nonetheless, if the Gemini theorem is correct, and it is also the case that human beings are capable of self-certainty, then it is difficult to avoid the conclusion that the Lucas-Penrose thesis must also be correct. In that sense the Gemini theorem, together with the empirical finding of self-certainty in humans, can be regarded as an independent confirmation of the thesis.

It is regrettable that none of those who have so far raised objections to the Gemini theorem have been prepared to submit their objections to peer-reviewed publication. This article is, in large part, an attempt to stimulate just such a discussion in the peer-reviewed literature, which is surely where it should be taking place. There is no doubt that the consequences of the Gemini theorem will be deeply unattractive to many, if not most, cognitive scientists and philosophers of mind. A temptation obviously exists for such individuals to rationalize their dislike of these consequences in terms of spurious and poorly argued objections which are never submitted to peer-reviewed publication. I therefore recommend all philosophers and cognitive scientists who have such objections to submit them for publication, so that they can be subjected to proper peer review and analysis.

In my work so far I have implicitly taken it for granted that human beings are capable of certainty. One may choose to question this assumption; it is however empirically testable. If self-certainty is indeed possible in human beings, then the Gemini theorem predicts that a violation of energy conservation (or $\chi$ effect) should in principle be detectable in human brains. Such a violation clearly has implications for any physical theory. Furthermore the Gemini theorem explicitly requires that such a violation cannot, even in principle, have a physical explanation. On the face of things this might seem to be bad news for physicists. However, I do not think that things are actually quite so bad. In fact, $\chi$ has some helpful properties. Firstly, even if we cannot say *how it works*, we *can* say exactly what it does. We can also say exactly when it will do it. In this respect, it resembles Turing's notion of an oracle, and can be treated in a similar way.

## References

Bojadziev, D. (2000). Is consciousness not a computational property? *Informatica,* 24, 75-77.

Caplain, G. (1995). Is consciousness a computational property? *Informatica,* 19, 615-619.

Caplain, G. (2000). Is consciousness not a computational property? - Reply to Bojadziev. *Informatica,* 24, 79-81.

Dunlop, C. E. M. (2000). Review of M. Gams, M. Paprzycki, and X. Wu (Eds) Mind Versus Computer: Were Dreyfus and Winograd Right? *Minds and Machines: Journal for Artificial Intelligence, Philosophy, and Cognitive Science* 10 (2), 289-296.

Parsons, T. (1996). What is an Argument?. *Journal of Philosophy,* 93 (4), 164-185. preprint available at:
http://philosophy.ucla.edu/wp-content/uploads/2016/10/WhatIsArg.pdf

Reason, C. M. (2016). Consciousness is Not a Physically Provable Property. *Journal of Mind and Behavior,* 37 (1), 31-46.

Reason, C. M. (2018). A Theoretical Solution of the Mind-Body Problem: An Operationalized Proof that no Purely Physical System Can Exhibit all the Properties of Human Consciousness. (*In preparation*) -- preprint available at: http://arxiv.org/abs/1706.04192

Shoemaker, S. (1990). First-Person Access. *Philosophical Perspectives,* 4, 187-214.